\documentclass[conference]{IEEEtran}

\usepackage{footnote}
\usepackage{booktabs} 
\usepackage{multirow}
\usepackage{url}
\usepackage{amsthm}
\usepackage{color}
\usepackage[cmex10]{amsmath}
\usepackage{listings}
\usepackage{graphicx}
\usepackage{xspace}
\usepackage[font=small,labelfont=bf]{caption}
\usepackage[colorlinks,allcolors=blue]{hyperref}
\usepackage{amssymb}
\title{OAT: Attesting Operation Integrity of \\ Embedded Devices}

\newcommand{\propname}{{Operation Execution Integrity}\xspace}
\newcommand{\propshortname}{{OEI}\xspace}
\newcommand{\toolname}{{OAT}\xspace}
\newcommand{\datacheckname}{{Value-based Define-Use Check}\xspace}

\newcommand{\cvi}{{CVI}\xspace}
\newcommand{\cvifull}{{Critical Variable Integrity}\xspace}

\newcommand{\point}[1]{\vspace{0.1in}\par\noindent{\bf #1:}}

\newcommand{\ie}{\textit{i.e.,}\xspace}
\newcommand*{\rom}[1]{\textit{\expandafter\romannumeral #1}}
\newcommand{\us}{\,$\mu$s\xspace}

\theoremstyle{definition}
\newtheorem{defn}{Definition}
\newenvironment{defi}
  {\pushQED{\qed}\defn}
  {\popQED\defn}

\def\zc{0}

\ifx\zc\undefined
	
\else
	
\fi

\newtheorem{theorem}{Theorem}
\newtheorem{definition}[theorem]{Definition}
\newtheorem{lemma}[theorem]{Lemma}

\definecolor{mygreen}{rgb}{0,0.6,0}
\definecolor{mygray}{rgb}{0.5,0.5,0.5}
\definecolor{mymauve}{rgb}{0.58,0,0.82}

\author{
   \IEEEauthorblockN{Zhichuang Sun}
   \IEEEauthorblockA{Northeastern University}
   \and
   \IEEEauthorblockN{Bo Feng}
   \IEEEauthorblockA{Northeastern University} 
   \and
   \IEEEauthorblockN{Long Lu}
   \IEEEauthorblockA{Northeastern University} 
   \and
   \IEEEauthorblockN{Somesh Jha}
   \IEEEauthorblockA{University of Wisconsin} 
}

\begin{document}
\maketitle

\begin{abstract}
Due to the wide adoption of IoT/CPS systems, embedded devices (IoT frontends)
become increasingly connected and mission-critical, which in turn has attracted
advanced attacks (e.g., control-flow hijacks and data-only attacks).
Unfortunately, IoT backends (e.g., remote controllers or in-cloud services) are
unable to detect if such attacks have happened while receiving data, service
requests, or operation status from IoT devices (remotely deployed embedded
devices). As a result, currently, IoT backends are forced to blindly trust the IoT
devices that they interact with.

To fill this void, we first formulate a new security property for embedded
devices, called {\it ``\propname''} or {\it \propshortname}. We then design and
build a system, \toolname, that enables remote \propshortname attestation
for ARM-based bare-metal embedded devices. 
Our formulation of \propshortname captures the integrity of both control flow
and critical data involved in an operation execution. Therefore, satisfying
\propshortname entails that an operation execution is free of unexpected control
and data manipulations, which existing attestation methods cannot check. Our
design of \toolname strikes a balance between prover's constraints (embedded
devices' limited computing power and storage) and verifier's requirements
(complete verifiability and forensic assistance). \toolname uses a new
control-flow measurement scheme, which enables lightweight and space-efficient
collection of measurements (97\% space reduction from the trace-based approach). \toolname
performs the remote control-flow verification through abstract execution, which
is fast and deterministic. \toolname also features lightweight integrity
checking for critical data (74\% less instrumentation needed than previous
work). Our security analysis shows that \toolname allows remote verifiers or
IoT backends to detect both control-flow hijacks and data-only attacks that
affect the execution of operations on IoT devices. In our evaluation using real
embedded programs, \toolname incurs a runtime overhead of 2.7\%.

\end{abstract}

\section{Introduction}
\label{sec:intro}

{\let\thefootnote\relax\footnotetext{
\kern -3pt
\hrule width 2in
\kern 2.6pt
To appear in the IEEE Symposium on Security \& Privacy, May 2020.
This work was supported by the National Science
Foundation (Grant\#: CNS-1748334), the Office of Naval Research (Grant\#:
N00014-17-1-2227), and the Army Research Office (Grant\#: W911NF-17-1-0039).
}}

Internet-of-Things (IoT) and Cyber-Physical Systems (CPS) are being rapidly
deployed in smart homes, automated factories, intelligent cities, and more. As a
result, embedded devices, playing the central roles as sensors, actuators, or
edge-computing nodes in IoT systems, are becoming attractive targets for cyber
attacks. Unlike computers, attacks on embedded devices can cause not only
software failures or data breaches but also physical damage. Moreover, a
compromised device can trick or manipulate the IoT backend (e.g., remote
controllers or in-cloud services): hijacking operations and
forging data.

Unfortunately, today's IoT backends cannot protect themselves from
manipulations by compromised IoT devices. This is due to the lack of a technique
for remotely verifying if an operation performed by an IoT device has been
disrupted, or any critical data has been corrupted while being processed on the
device. As a result, IoT backends are forced to blindly trust remote devices
for faithfully performing assigned operations and providing genuine data. Our
work aims to make this trust verifiable, and therefore, prevent compromised IoT
devices from deceiving or manipulating the IoT backend.

We take the general approach of remote attestation, which allows a device to
prove its integrity (with regard to certain security properties) to a remote
verifier. Although a large body of works on remote attestation
exists~\cite{Seshadri2004,Arvind04softAt,Haldar04semantic,
Seshadri2005,Castelluccia2009, Kil09ReDAS}, they have different goals from ours.
Moreover, most of them are focused on verifying basic security properties, such as
static code text integrity, and therefore cannot capture the advanced attacks
that are becoming mainstream recently. For example, return-oriented programming
(ROP) and data-only attacks are easy to launch on embedded devices, as
demonstrated on vulnerable industrial robot controllers~\cite{quarta17sp}. 

C-FLAT~\cite{Abera2016} took the first step towards control-flow integrity (CFI) attestation. 
But a major limitation of C-FLAT is the non-deterministic verifiability of its 
control-flow hashes (i.e., a given hash may not be verifiable due to the program 
path explosion issue). Moreover, C-FLAT does not check data integrity 
(i.e., data-only attacks are not covered). 

\vspace{1em}
In this paper, we introduce the first attestation method that captures both
control-flow and data-only attacks on embedded devices. Using this method, IoT
backends can now verify if a remote device is trustworthy when it claims it has
performed an operation, sent in a service request, or transported back data from
the field. In addition, unlike traditional attestation methods, which only output
a binary result, our method allows verifiers to reconstruct attack execution
traces for postmortem analysis. 

Our attestation is based on a new security property that we formulated, called
{\em \propname} (\propshortname), which combines both the control-flow integrity
and critical data integrity of an ``{\em operation}'' (i.e., a self-contained
task or logic). An operation satisfies \propshortname if and only if the
operation was performed without its control flow altered or its critical data
corrupted during the execution. For an operation to be attested, the IoT device
(i.e., prover) sends an unforgeable \propshortname ~{\em measurement} to the IoT
backend (i.e., verifier), along with any output from the operation. The
backend then checks the measurement to determine if \propshortname was
satisfied during the operation. The backend accepts the operation output from the device
only if the check passes (i.e., the received data or request is trustworthy).

\propshortname takes advantage of the ``operation-oriented'' design of embedded
programs: code is typically organized in logically independent operations, such
as moving a robotic arm, injecting a dose of medicine, sensing temperature, etc.
Rather than covering an entire program, \propshortname is focused on the
execution of individual operations (hence the name). This per-operation property
allows for on-demand and efficient attestation on embedded devices without
sacrificing security (\S\ref{sec:oeiDef}).

We design and implement \toolname (\underline{O}EI \underline{AT}tester), a
system that enables \propshortname attestation on ARM-based embedded devices. It
consists of a customized compiler for building attestation-enabled binaries, a
runtime measurement engine running on IoT devices, and a verification engine for
IoT backends. \toolname addresses two key challenges associated with OEI
attestation, or any remote attestation of control-flow and data integrity:  

\point{(1) Incomplete verification of control-flow integrity} Conventional
hash-based attestation~\cite{Abera2016} can only verify a (small) subset of program executions
 (i.e., incomplete verification of control-flow). It is because this
approach checks a given control-flow hash against a limited set of hashes
pre-computed from known-legitimate program runs. This static hash pre-computation can
never cover all possibilities due to program path explosions, even for small
programs. As a result, this attestation cannot verify control-flow hashes,
legitimate or not, outside of the pre-computed set. 

We design a hybrid control-flow attestation scheme for \toolname, which
combines hashes and compact execution traces. This scheme enables complete
control-flow verification as well as attack flow reconstruction, at the cost of
a mildly increased measurement size. Our attestation scheme is partly inspired
by the tracing-based CFI enforcement~\cite{ge2017griffin,van2015practical}. But
unlike previous work, which requires hardware tracing modules unavailable
on deployed or debugging-disabled embedded devices, 

our scheme uses its own software-based tracing
technique. Moreover, thanks to the combined use of hash and traces, \toolname's space overhead is only a tiny fraction (2.24\%) of tracing-based CFI's overhead. 
In addition, \toolname checks both
forward- and backward-edges. We discuss the details in \S\ref{sec:cf}.

\point{(2) Heavy data integrity checking} 
The existing data integrity
checkers~\cite{Castro2006,akritidis2008preventing,carr2017datashield} have to
instrument every memory-write instruction and sometimes memory-read instructions in a program.
The heavy and extensive instrumentation is needed because these checkers have to
decide during runtime, for every instrumented instruction, whether the instruction
is allowed to store/load data to/from the referenced address. We
call this {\em address-based checking}, which is too heavy for embedded
devices.  

\toolname uses a novel data integrity checking technique. First, it only covers
critical variables because not all program data is relevant to an operation to
be attested. Critical variables are those that may affect the outcome of an
operation. They are automatically detected by \toolname or annotated by
developers.  

Second, instead of address-based checking, our technique performs {\em
value-based checking}. It checks if the value of a critical variable at an
instrumented load instruction (i.e., use) remains the same as the value recorded
at the previous instrumented store instruction (i.e., define). It only
instruments the instructions that are supposed to access the critical variables,
rather than instrumenting all memory-accessing instructions as address-based
checkers would, even when only selected variables need checking. Our technique on
average requires 74\% fewer instrumentation than address-based
checking does. We call this ``\datacheckname'', which is discussed in \S\ref{sec:cvi}.

\vspace{1em}
\noindent
Using \toolname, IoT backends can now for the first time remotely verify
operations performed by IoT devices. Our security analysis
(\S\ref{sec:security_analysis}) shows that \toolname detects both control and
data manipulations that are undetectable by existing attestation methods for
embedded devices. Our performance evaluation (\S\ref{sec:eval}), based on real
embedded programs, shows that \toolname, on average, slows down program
execution by 2.73\% and increases the binary size by 13\%. 

In summary, our work makes the following contributions:
\begin{itemize}
\item We formulate a new security property, \propshortname, for IoT backends to attest the integrity of operations executed on remote IoT/embedded devices. It covers both control-flow and critical data integrity. 
\item We design a hybrid attestation scheme, which uses both hashes and execution traces to achieve complete control-flow verification while keeping the size of control-flow measurements acceptable for embedded devices. 
\item We present a light-weight variable integrity checking mechanism, which uses selective and value-based checking to keep the overhead low without sacrificing security. 
\item We design and build \toolname to realize \propshortname attestation on both the prover- and verifier-side. \toolname contains the compile-time, run-time, and verification-time components. 
\item We evaluate \toolname on five real-world embedded programs that cover broad use scenarios of IoT devices, demonstrating the practicality of \toolname in real-world cases.
\end{itemize}

\section{Background}
\label{sec:bg}

\subsection{Attacks on IoT Devices and Backends}

Embedded devices, essential for IoT, have been
increasingly targeted by powerful attacks. For instance, hackers have managed to
subvert different kinds of smart home gadgets, including connected
lights~\cite{ronen2017iot}, locks~\cite{ho2016smart}, etc. In industrial
systems, robot controllers~\cite{quarta17sp} and PLCs (Programmable Logic
Controller)~\cite{falliere2011w32} were exploited to perform unintended or
harmful operations. The same goes for connected cars~\cite{keenlab,nissan},
drones~\cite{greenberg2016hacker}, and medical devices~\cite{pacemaker,insulin}.
In addition, large-scale IoT deployments were compromised to form botnets via
password cracking~\cite{mirai}, and recently, vulnerability
exploits~\cite{reaper}.

Meanwhile, advanced attacks quickly emerged. 
Return-oriented Programming (ROP) was demonstrated to be
realistic on RISC~\cite{buchanan2008good}, and particularly
ARM~\cite{kornau2010return}, which is the common architecture for today's
embedded devices. Data-only attacks~\cite{chen2005non,hu2016data}
are not just applicable but well-suited for embedded devices~\cite{wurm2016security}, due to the
data-intensive or data-driven nature of IoT. 

Due to the poor security of today's embedded devices, IoT backends (e.g., remote
IoT controllers and in-cloud services) are recommended to operate under the
assumption that IoT devices in the field can be compromised and should not be
fully trusted ~\cite{stergiou2018secure}. However, in reality, IoT backends are often
helpless when deciding whether or to what extent it should trust an IoT device.
They may resort to the existing remote attestation techniques, but these
techniques are only effective at detecting the basic attacks (e.g., device or code
modification) while leaving advanced attacks undetected (e.g., ROP, data-only
attacks, etc.). As a result, IoT backends have no choice but to trust IoT
devices and assume they would faithfully execute commands and generate genuine data or requests.
This blind and unwarranted trust can subject IoT backends to deceptions and
manipulations. For example, a compromised robotic arm can drop a command yet
still report a success back to its controller; a compromised industrial syringe
can perform an unauthorized chemical injection, or change an authorized
injection volume, without the controller's knowledge. 

Our work enables IoT backends to reliably verify if an operation performed by a
device has suffered from control or data attacks. It solves an important open
problem that IoT backends currently have no means to determine if data,
results, or requests sent from (insecure) IoT devices are trustworthy. Moreover,
it allows backends to reconstruct attack control flows, which are valuable for
forensic analysis.

\subsection{ARM TrustZone}

Our system relies on ARM TrustZone to establish the TCB (Trusted Computing Base). 
TrustZone is a hardware
feature available on both Cortex-A processors (for mobile and high-end IoT
devices) and Cortex-M processors (for low-cost embedded systems). TrustZone
renders a so-called ``Secure World'', an isolated environment with tagged caches,
banked registers, and private memory for securely executing a stack of trusted
software, including a tiny OS and trusted applications (TA). In parallel runs
the so-called ``Normal World'', which contains the regular/untrusted software stack.
Code in the Normal World, called client applications (CA), can invoke TAs in the
Secure World. A typical use of TrustZone involves a CA requesting a sensitive
service from a corresponding TA, such as signing or securely storing a piece of
data. In addition to executing critical code, TrustZone can also be used for
mediating peripheral access from the Normal World. 

\toolname measurement engine runs as a TA in the Secure World. Its code and
data, including collected measurements, are naturally protected. TrustZone
allows provisions of per-device private keys and certificates, which enable
straightforward authentication and signed/encrypted communication between a
measurement engine (i.e., prover) and a remote verifier. During an active
attestation phase, the instrumented code in the Normal World reports raw
measurements to the Secure World, where the raw measurements are processed and
signed. The final report (aka. measurement blob) along with the signature is
handed back to the Normal World agent, which sends the report to the remote
verifier. On our target platforms (i.e., ARM-based bare-metal embedded devices),
TrustZone is the only feasible TCB option that can support basic remote
attestation operations. Our evaluation (\S\ref{sec:eval}) shows that the end-to-end overhead of our
system is acceptable, thanks to our efficient attestation scheme.

\section{Design Overview}
\label{sec:design}

\subsection{Example: A Vulnerable Robotic Arm}
\label{sec:runningexample} Before we discuss \propshortname and the attestation,
we first present a simple example to demonstrate the problem. The vulnerabilities shown in
this example were drawn from real embedded programs. This example is also
referenced in a later discussion on how
\propshortname attestation can be easily applied to existing code.

In this example of a vulnerable robotic arm (Listing~\ref{lst:examplecode}), the
function {\tt main\_looper} first retrieves an operation command from a remote controller and stores it
in {\tt cmd} (Line 11). The looper then reads a peripheral sensor (Line 12),
which indicates if the arm is ready to perform a new operation. If ready, the
looper finds the operation function specified by {\tt cmd->op} (Line 15) and
calls the function with the parameters supplied in {\tt cmd->param} (Line 16).
One such function (Line 25) moves the arm to a given position.

\begin{lstlisting}[caption={An example of a control loop in a robotic arm},
  captionpos=b,label={lst:examplecode},language=C,basicstyle=\footnotesize\ttfamily,float=tp,xleftmargin=.02\textwidth, xrightmargin=.02\textwidth]
// Simplified control loop for robotic arms
void main_looper() {
  // Command from remote controller
  cmd_t cmd; 
  // Pointer to operation function
  int (*op_func)(int, char*);
  // Input from peripheral sensor
  char peripheral_input[128];
  int st = 0;
  while(1) {
    if (read_command(&cmd)) {
      st = get_input(peripheral_input); //<@\textcolor{red}{BUGGY!}@>
      if (status_OK(st)) {
        // perform the operation
        op_func = get_op_func(cmd->op);        
        (*op_func)(cmd->p_size, cmd->param);
      }
    }
    usleep(LOOPER_SLEEP_TIMER); 
  }
  return;
}
...
// The operation that moves the robotic arm
int op_move_arm (int, char*) {...}
...
\end{lstlisting}

We introduce an attacker whose goal is to manipulate the operation of the
robotic arm without being detected by the remote controller who uses the
existing attestation methods. For simplicity, we assume the attacker has
tampered with the sensor and uses it to feed exploit input to the robotic arm.
This is realistic given that such external peripherals are difficult to
authenticate and protect. 
The target of the attacker is Function {\tt get\_input} (Line 12), which
contains a buffer overrun. The vulnerability allows malformatted input to be
copied beyond the destination buffer ({\tt peripheral\_input}) into the
subsequent stack variables (e.g., {\tt cmd}).

By crafting input via the compromised sensor, the attacker can launch either
control hijacking or data-only attacks on the robotic arm. To hijack the
control, the attacker overwrites both {\tt cmd->param} and {\tt cmd->op} as a
result of the buffer overrun exploit, which leads to the execution of an
arbitrary operation. To mount a data-only attack, the attacker only needs to change {\tt
cmd->param} while leaving {\tt cmd->op} intact, which turns the authorized
operation harmful.

Though seemingly simple, such control and data manipulations on embedded devices
are realistic and can cause severe consequences in the physical world. More
importantly, existing remote attestation methods cannot detect such attacks
because most of them are focused on static code integrity and none addresses
dynamic data integrity. Therefore, the remote controller is unable to find that
the robotic arm is compromised and did not perform the operation as commanded.
Moreover, after receiving an (unverifiable) operation-success message from the
compromised robotic arm, the controller may command other robotic arms to
perform follow-up operations, which can cause further damage. This example
illustrates the need for a new form of attestation that allows IoT backends to
remotely check the integrity of operations performed by IoT devices.

\subsection{\propname (\propshortname)}
\label{sec:oeiDef}
We propose ``\propname'' (or \propshortname) as a security property for embedded
devices. By verifying \propshortname, a remote verifier can quickly check if an 
{\em execution of an operation} suffered from {\bf control-flow hijack} or
experienced {\bf critical data corruption}.
We formulate \propshortname with two goals in mind: (\rom{1}) enabling remote
detection of aforementioned attacks; (\rom{2}) demonstrating the feasibility of
an operation-oriented attestation that can detect to both control and critical
data manipulations on embedded devices. Next, we formally define \propshortname
and provide its rationale. 

To avoid ambiguity, we informally define an {\bf operation} to be a logically
independent task performed by an embedded device. To declare an operation for
attestation, programmers need to identify the entry and exit points of the
operation in their code. For simplicity of discussion, we assume that every
operation is implemented in a way that it has a single pair of entry and exit,
$\langle Op_{entry}, Op_{exit} \rangle$ where $Op_{entry}$ dominates $Op_{exit}$
and $Op_{exit}$ post-dominates $Op_{entry}$ in the control flow graph. We do not
pose any restriction on what code can an operation include or invoke, except
that an operation cannot contain another operation.

Let $P=\{Op_1, Op_2, ... Op_n\}$ be an embedded program, composed of $n$
operations, denoted as $Op_i$. $CFG(Op_i)$ is the statically constructed CFG (control
flow graph) for $Op_i$ (i.e., the CFG's root is $Op_i$'s entry point). Let $CV$ be
the set of critical variables in $P$ (i.e., variables affecting the execution of the
operations). 
$D(v)$ and $U(v)$ are the def- and use-sites
of a critical variable $v$: a set of statements where $v$ is {\em defined} or {\em used}.
$V_b(v,s)$ and $V_a(v,s)$ are the values of variable $v$ immediately {\em
before} and {\em after} statement $s$ executes. 

\begin{defi}
  \label{def:OEI}
  {\bf \propshortname:} for an operation $Op_i$, its execution satisfies
  OEI $\iff$ ~\textcircled{1} the control flow complies with $CFG(Op_i)$ and maintains backward-edge integrity, and ~\textcircled{2} during the execution, 
  $\forall cv \in CV$, the value of $cv$ reaching each use-site is the same as the value of $cv$ leaving the previous def-site, written as $V_b(cv,u)=V_a(cv,d)$, where $d \in D(cv)$ is the
  last define of $cv$ prior to the use of $cv$ at $u \in U(cv)$. 
\end{defi}

OEI entails that the control-flows and critical data involved in an operation must not be tampered with.

\point{Operation-scoped CFI (\S\ref{sec:cf})} OEI's control-flow requirement (\textcircled{1} in Def.~\ref{def:OEI}) is not a simple adoption of CFI to embedded
devices, which would incur impractical time and space overhead on those
constrained devices. Our operation-scoped CFI takes advantage of the
operation-oriented design of embedded programs. It applies to executions of
individual operations, which represent a much smaller attestation scope
than a whole program and allow for on-demand attestation. It also implies backward-edge integrity (i.e., returns not hijacked).

\point{\cvifull (\S\ref{sec:cvi})} We call the second requirement of OEI
(\textcircled{2} in Def.~\ref{def:OEI}) \cvifull, or \cvi. It dictates that the
{\em values} of critical variables must obey {\em define-use consistency}.
Compared with other data integrity
checkers~\cite{Castro2006,akritidis2008preventing,carr2017datashield} \cvi is
different in two ways. First, \cvi only concerns critical variables, rather than
all program data. Second, \cvi uses value-based checking, instead of 
address-based checking, to significantly reduce code instrumentation and runtime
overhead.
\cvi applies to the entire program execution and is not
scoped by attested operations. We provide a method to assist developers to
automatically identify critical variables. 
We define critical variables and explain our value-based check in \S\ref{sec:cvi}.

\point{Secure \& Optimized Combination} OEI combines \cvi and operation-scoped CFI. These
two sub-properties mutually complementary. Without \cvi,
CFI alone cannot detect data-only attacks or control-flow hijacks that do not
violate CFG (e.g., \cite{carlini2015control}). Without CFI, \cvi can be bypassed
(e.g., by ROP).
On the other hand, this combination yields better performance than independently
performing control-flow and data-flow checks. This optimized combination allows
for the detection of both control-flow and data-only attacks without enforcing
full CFI and DFI. It is suited for embedded devices.

\point{Operation Verifiability}
\propshortname caters to IoT's unique security needs. One defining
characteristic of IoT devices is their frequent inter-operations with peers or cloud
services. When a device finishes an operation, the operation initiator may wish
to verify if the operation was executed as expected without interference. For
example, a remote controller needs to verify if a robotic arm has performed a
movement as instructed without experiencing any control or data manipulations.
OEI attestation answers to such
security needs of IoT, which are currently not addressed.

\subsection{\toolname Overview}

\begin{figure}
    \centering
      \includegraphics[width=8cm]{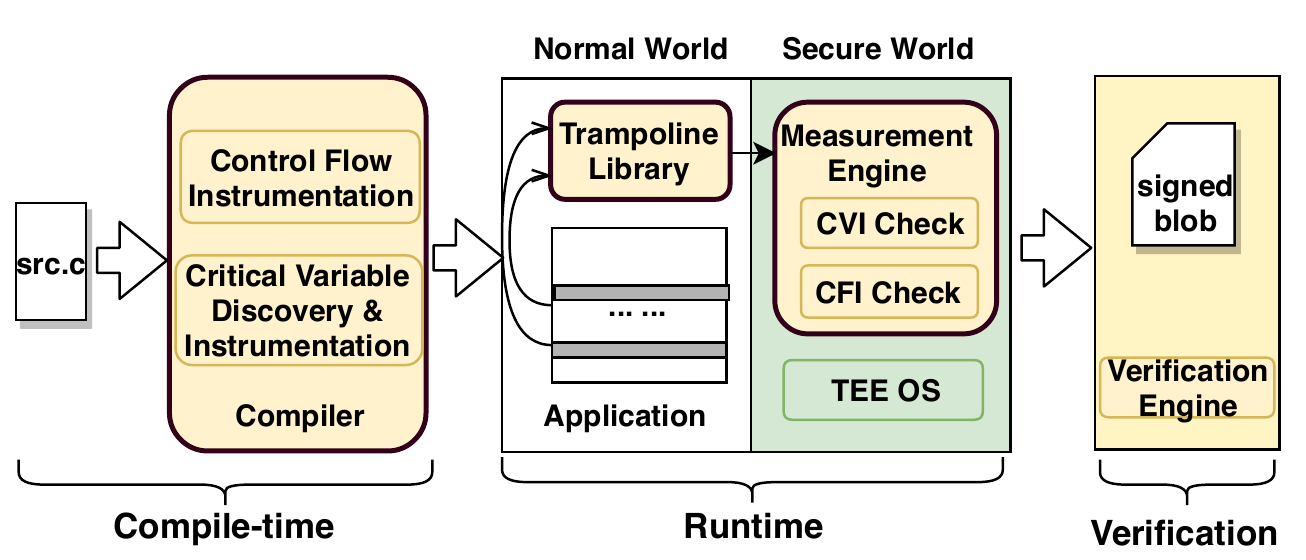}
      \caption{The Workflow of OAT, whose components (colored in
        yellow) include the compiler, the trampoline library, the
        measurement engine in the Secure World (shown in
        green), and the remote verification engine.}
      \label{fig:designoverview}
    \end{figure}

We build \toolname, a system to realize \underline{O}EI \underline{AT}testation
on ARM-based bare-metal embedded devices (i.e., no standalone OS in the Normal
World). 
\toolname consists of: (\rom{1}) a customized {\bf compiler} built on LLVM;
(\rom{2}) a {\bf trampoline} library linked to attestation-enabled programs;
(\rom{3}) a runtime {\bf measurement engine}; (\rom{4}) a remote/offline {\bf
verification engine} (Figure~\ref{fig:designoverview}). For a given embedded
program, the compiler identifies the critical variables and instruments the code
for measurement collection.  At runtime, the measurement engine processes the
instrumented control-flow and data events and produces a proof or measurement,
which the remote verifier checks and determines if the operation execution meets
OEI. \toolname relies on ARM TrustZone to provide the trusted execution
environment for the measurement engine.

We design a hybrid attestation scheme that overcomes two challenges
associated with CFI and data integrity verifications. First, remotely attesting
CFI is more challenging than performing local CFI enforcement because remote
verifiers can only rely on limited after-fact measurements to make the
determination whereas local enforcers simply use the readily available unlimited
runtime information. Furthermore, remote verifiers cannot always determine
whether a hash-based control-flow measurement is legitimate or not because the
verifiability is limited to those hashes pre-computed from known/traversed code
paths. 

Second, checking the integrity of dynamic data is known for its high overhead.
Existing checkers~\cite{Castro2006,akritidis2008preventing,carr2017datashield},
mostly address-based, instrument every memory-accessing instruction in a
program, and during runtime, check if an instrumented instruction should be
allowed to access the referenced address, based on a statically constructed
access table. Even when the integrity checking is only applied to selected
variables, address-based checkers would still need to instrument and check all
memory-accessing instructions to ensure no unintended instructions can write to
the addresses of the critical variables.

Our hybrid attestation scheme achieves complete verifiability while maintaining
acceptable performance on embedded devices. For CFI attestation, \toolname's
measurements contain compact control-flow transfer traces for forward-edges and
fixed-length hashes for backward-edges. This combination allows remote
verifiers to quickly and deterministically reconstruct control flows. It also
yields size-optimized measurements.  
For \cvi, \toolname performs local verification rather than remote attestation.
By doing so, \toolname avoids sending a large amount of data (e.g., critical
values at def- and use-sites) to remote verifiers. Sending such data would be
costly for IoT devices and undesirable when privacy is concerned.

To use OEI attestation, programmers declare the to-be-attested
operations in their code by using two intuitive compiler directives: {\tt
\#oei\_attest\_begin} and {\tt \#oei\_attest\_end}. They may also annotate
critical variables of their choice via a GCC-style attribute. 
For example, to enable
\propshortname attestation in Listing~\ref{lst:examplecode}, a programmer only
needs to change Line 4, 8, and 25: 

\vspace{1em}
\begin{lstlisting}[language=C,basicstyle=\footnotesize\ttfamily,firstnumber=4,xleftmargin=.02\textwidth, xrightmargin=.02\textwidth]
cmd_t __attribute__((annotate("critical"))) cmd; 
\end{lstlisting}
\vspace{-2em}
\begin{lstlisting}[language=C,basicstyle=\footnotesize\ttfamily,firstnumber=8,xleftmargin=.02\textwidth, xrightmargin=.02\textwidth]
int __attribute__((annotate("critical"))) peripheral_input = 0;
\end{lstlisting}
\vspace{-2em}
\begin{lstlisting}[language=C,basicstyle=\footnotesize\ttfamily,firstnumber=25,xleftmargin=.02\textwidth, xrightmargin=.02\textwidth]
int op_move_arm (int, char*) {#oei_attest_begin ... #oei_attest_end}
\end{lstlisting}
\vspace{-1em}

For simplicity, our current design requires that a pair of {\tt
\#oei\_attest\_begin} and {\tt \#oei\_attest\_end} is used in the same function
(i.e., an operation enters and exits in the same call stack frame) and the {\tt
\#oei\_attest\_end} always dominates the {\tt \#oei\_attest\_begin} (i.e., an
operation always exits). Operations cannot be nested. These requirements are
checked by our compiler. Developers are advised to keep the scope of an
operation minimal and the logic self-contained, which is not difficult because
most embedded programs are already written in an operation-oriented fashion. 

As shown in Figure~\ref{fig:designoverview}, during compilation, the customized
compiler instruments a set of control flow transfers inside each to-be-attested
operation. The compiler automatically annotates control-dependent variables as
critical (e.g., condition variables). It also performs a global data dependency
analysis on both automatically and manually annotated critical variables so that
their dependencies are also treated as critical and subject to \cvi checks. 
At runtime, the instrumented control flow transfers and critical variable
define/use events trigger the corresponding trampolines, which pass the
information needed for CFI or \cvi verification to the measurement engine in the
Secure World protected by TrustZone. Finally, the signed attestation blob (i.e.,
measurements) is sent to the remote verification engine (e.g., IoT
backend) along with the output from the attested operation. 
We discuss OAT design details in ~\S\ref{sec:cf} and ~\S\ref{sec:cvi}.

\subsection{Threat Model}
\label{sec:threatModel}

We trust code running inside the Secure World (e.g., the measurement engine) and
assume that attackers cannot break the TrustZone protection. We also trust our
compiler and the trampoline code. We assume that attackers {\em cannot inject
code in the Normal World or tamper with the instrumented code or the trampoline
library}. This can be enforced using code integrity protection methods for
embedded devices~\cite{clements2017protecting,kimsecuring}, which are orthogonal
to the focus of this paper, namely \propshortname attestation. Due to the
absence of a standalone OS on bare-metal embedded devices, all code in the
Normal World runs at the same privilege level (i.e., no higher privileged code
exists or needs to be trusted).

On the other hand, we anticipate the following attacks in the Normal World,
which previous attestation methods cannot detect. First, attackers may exploit
vulnerabilities to launch ROP (return-oriented programming) and DOP
(data-oriented programming) attacks. As a result, the control flow and the
critical data of an embedded program can be compromised. Second, attackers may
abuse unprotected interfaces of an embedded program and force the device to
perform unintended or unauthorized operations. Our system is designed to detect
these attacks by means of attestation. We present a security analysis on our
system in~\S\ref{sec:security}.

\section{Operation-scoped Control-flow Attestation}
\label{sec:cf}

Inspired by the operation-oriented nature of embedded programs, we attest CFI at
the operation level, which avoids always-on measurement collection and
whole-program instrumentation. It is lightweight and suitable for embedded
devices. 

Moreover, our attestation provides deterministic verifiability. It avoids the
problem of unverifiable control-flow hashes, caused by code path explosions, as
purely hash-based attestation faces~\cite{Abera2016}. The deterministic
verifiability is achieved via a new hybrid measurement scheme, which uses a
compact trace for recording forward edges and a hash for backward edges. This
scheme resembles the hardware tracing-based
CFI~\cite{ge2017griffin,van2015practical}. But it has two major distinctions.

First, previous tracing-based CFI requires hardware components that are not
available on deployed IoT or embedded devices\footnote{Although recent MCUs support tracing, this optional feature is meant for debugging and usually unavailable on for-release devices due to additional hardware cost.}. Our control flow traces are generated
purely using the software. Second, our trace is much shorter and more compact partly
because it only records forward edges (i.e., backward edges or returns happen
very frequently and thus would lead to overly long traces that embedded
devices cannot store). 
By hashing the backward edges, rather than recording them
in the trace, our scheme reduces the trace size by 97\% (\S\ref{sec:evalreal}). Furthermore, 
combining trace and hash makes verification deterministic and free of
path explosions. Verifiers no longer need to
pre-compute or search for all possible code paths; they simply follow the forward-edge trace to reconstruct the actual execution path, and in the end, check the resulting backward-edge hash (more details later).

\point{Instrumented Control Flow Events}
\toolname compiler instruments the code in each attestation-enabled operation to
collect runtime measurements.
We limit this instrumentation to a minimum set
of the control flow transfers that need to be recorded/encoded in the
measurement for deterministic verification. This minimum set, $S_{min}$, is constructed as follows. Consider
all three types of transfers possible on a CFG: direct call/jump, conditional branch, and indirect transfers (indirect call/jump and return). 
Only the last two types need to be measured and are
included in $S_{min}$. This is because knowing where each branch and indirect
transfer went during a code execution is both sufficient and necessary for
statically finding the exact operation execution path on the CFG. 
Direct calls and jumps are not included in
$S_{min}$ (i.e., no need for instrumentation) because their destinations are unique
and statically determinate.

\begin{figure*}
    \centering
    \includegraphics[scale=0.65]{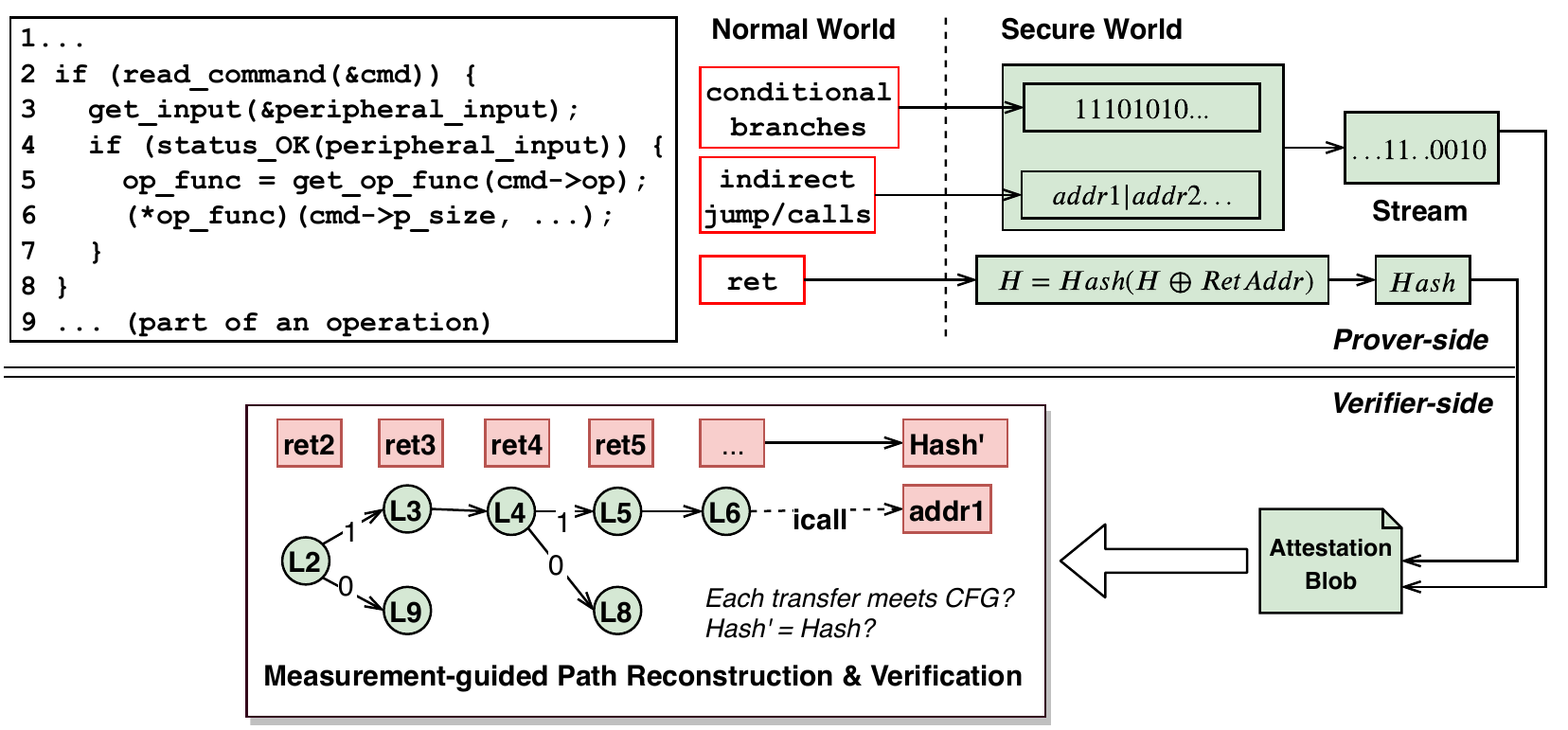}
    \caption{Operation-scoped control-flow attestation. By measuring the control-flow of an executing operation, the measurement engine produces a control-flow proof that consists of a {\em trace} (proving forward edges including branches and indirect transfers) and a {\em hash} (proving backward edges or returns). This measurement scheme allows a remote verifier to statically and deterministically reconstruct the code path of the operation execution and perform full CFI verification. 
    }
    \label{fig:control_flow_check}
\end{figure*}

The instrumentation code simply calls the trampoline function corresponding to
the type of the instrumented control-flow transfer, reporting the destination of
the transfer. For a branch, its destination is one bit:  {\em
taken} (1) or {\em non-taken} (0). For an indirect transfer, its destination is the memory address.
The trampoline functions are thin wrappers of the world-switching routine. When
called, they initiate the switch to the Secure World and pass the destination information to
the measurement engine.

\point{Measurement Scheme}
The measurement engine maintains two types of measurements for control-flow attestation:
a {\bf trace} and a {\bf hash}. The trace is used for recording forward-edge
control flow transfers (branches and indirect calls/jumps). The hash is for
encoding backward-edge transfers (returns). 
The measurement engine updates the measurement trace or hash respectively, as
shown in the upper half of Fig.~\ref{fig:control_flow_check}. For each branch,
it adds the taken/non-taken bit to the trace. For each indirect call/jump, it
adds the destination address to the trace. Note that code addresses do not
change across different firmware runs (including verification runs) because
embedded firmware is always loaded at the pre-specified base address in memory.
When dynamic loading or ASLR becomes available in embedded firmware, the
measurement engine will need to record the firmware base address in the
attestation blob, in order for the verifier to construct the same code layout in
memory and check the trace. For each return, the measurement engine encodes the
return address to the hash: $H=Hash(H \oplus RetAddr)$. Here we use the symbol
$\oplus$ to represent a binary operation. In our implementation, we use
concatenation\footnote{See our formal definition and proof of the verification
scheme in Appendix  ~\S\ref{sec:prf}}. The trace and the hash together form
the attestation blob, which serves as the proof of the control flow of an
executed operation.

In our design, we chose the cryptographic hash
function BLAKE-2s\footnote{https://blake2.net/} as the $Hash$ function
for its high speed, security, and simplicity.
BLAKE-2s has a block size of 64 Bytes and an output size of 32 Bytes. We
present the collision analysis as part of the security analysis in \S~\ref{sec:security}.  

The reason why we use two forms of measurements, namely trace and hash,
is two-fold. First, a forward-edge trace allows for reconstruction and easy verification of
recorded control-flow transfers. 
Second, a hash has fixed length and does not grow whereas a trace grows as
 the execution proceeds. However, control-flow hashes by themselves are not
 always verifiable due to the impossibility of pre-computing all possible code
 paths for a program and their hashes. Our measurement scheme uses the
 trace and hash in tandem to combine their strengths while avoiding their
 individual weaknesses, in terms of the ease of verification and space
 efficiency. 

Recording the forward-edge trace  is
necessary for code path reconstruction. These events are either very compact (1
bit for each branch) or infrequent (indirect calls/jumps are less common than
direct calls/jumps). Therefore, they do not bloat the trace. On the other hand,
we encode return addresses in the hash because they are not needed during the
path reconstruction phase (i.e., only needed for checking backward-edge CFI
after a path has been constructed). Plus, returns happen fairly frequently and,
if recorded in the trace, would consume too much space. 

A possible (yet to implement) optimization would be to record the event type
information in a separate sequence and then use this information to enable early
detection of control flow divergences during verification (i.e., when a type
mismatch is detected, the verifier can conclude that the current stream is
invalid and terminate the verification early). For three types of events (\ie
conditional branch, indirect branch and return), two bits would be enough for
identifying and recording each type. This optimization can speed up the
verification process at the cost of using extra storage. However, given the fact
that RAM and flash storage on embedded devices are fairly limited and the
verification process does not affect the runtime performance, we decided not to
implement the early termination optimization for the current prototype. 

The measurements are stored in a buffer allocated in the Secure World. Although rare, 
this buffer can run out if an operation execution is very long and yields a measurement 
trace longer than the buffer. When this happens, the measurement engine signs the current 
measurements, temporarily stores it in the flash storage, and frees up the buffer for 
subsequent measurements. At the end of the operation, the measurement engine sends all 
measurements to the remote verifier.

\point{Measurement Verification}
Given a control-flow proof for an operation execution (i.e., a trace and a
hash), generated and signed by the measurement engine, the verification engine
statically reconstructs the code path on the CFG of the operation, as shown in the lower half of Fig.~\ref{fig:control_flow_check}.   
Starting from
the root basic block, or the entry point, the verifier abstractly executes the
operation on the CFG by following the forward-edge trace. During the abstract execution,
the verifier maintains a simulated call stack to keep track of the return addresses
and computes/updates the hash in the same way as the measurement engine does
during runtime.

Specifically, 
when the abstract execution encounters a control-flow diverging point
(i.e., more than one edge on the CFG can be followed)
the verifier takes the next available element out of the trace and consults it
for direction: either a taken/non-taken bit for a branch or an address for an
indirect call/jump. The verifier also performs a CFI check in case of an
address. A control flow violation is found when: ~\textcircled{1} the CFI check fails;
~\textcircled{2} a mismatch is observed between a basic block and the corresponding trace
element (e.g.,  the current basic block ends with an indirect call but the next
available element indicates a branch); or ~\textcircled{3} after the abstract execution, 
the verifier-computed hash does not match the reported hash. 

All indirect call/jump violating CFI trigger ~\textcircled{1}. ~\textcircled{2}
can happen because when ROP happened during the actual execution, the trace did
not return to the call site whereas the abstract execution always returns
correctly (cued by the simulated call stack).  This mismatch leads to early
detection of ROP (i.e., no need to wait till ~\textcircled{3}).
~\textcircled{3} signals that one or more return addresses (or backward control
flow transfers) were corrupted during runtime. Note that not all ROP trigger
~\textcircled{2} but all ROP trigger ~\textcircled{3}.

\section{\cvifull}
\label{sec:cvi}

Data-only attacks, including data-oriented programming, are capable of
manipulating program behavior and internal logic without injecting code or
violating CFI~\cite{chen2005non,hu2016data}. 
For example, as shown in Listing~\ref{lst:examplecode}, a simple buffer overflow
can make the robotic arm perform attacker-specified operations without breaking
CFI. Unfortunately, existing attestation methods cannot detect data-only
attacks. 

We formulate \cvifull (\cvi) as a sub-property of \propshortname to detect
data-only attacks on embedded devices. \cvi is selective and concerns only data
that attacks target: (\rom{1}) {\bf control-dependent data}, such as conditional
variables, whose value directly determines the code path of a particular program
execution; (\rom{2}) {\bf semantically critical data}, whose value, though not
directly affecting code execution paths, influences the outcome of an operation,
such as {\tt cmd} in Listing~\ref{lst:examplecode}. \cvi
is not scoped by attested operations due to external data dependencies.

\cvi is different from previous works on data integrity
check~\cite{Castro2006,akritidis2008preventing,carr2017datashield}, which
require heavy instrumentation and is unsuitable for embedded devices.
For example, DFI~\cite{Castro2006} uses a whitelist to record which
instruction is allowed to access which memory address. It instruments all
memory-accessing instructions to perform the check during runtime. 
DataShield~\cite{carr2017datashield} reserves a special memory region
for critical variables to facilitate checks. Even though it concerns only
selected data, DataShield still needs to instrument all memory-write
instructions to prevent illegal access to the reserved memory region.
In general, previous works on program data integrity share the same fundamental
approach, which we call {\em address-based checking}. They need to check every
memory-write instruction and determines if it should be allowed to write to the
{\em referenced address}.   

In contrast, \cvi takes a new approach to data integrity checking, called
\datacheckname. It checks if the value of a critical variable at an instrumented
load instruction (i.e., use) remains the same as the value recorded at the
previous instrumented store instruction (i.e., define). At an instrumented store
instruction, \cvi makes a copy of the value in a secure region guarded by
TrustZone. At an instrumented load instruction, \cvi compares the value read
from memory with the copy recorded in TrustZone. Any data corruption causes a
value mismatch and therefore a \cvi breach. \cvi only needs to instrument the
instructions that are supposed to read/write the critical variables whereas
address-based checkers have to instrument all memory-write instructions even if
only selected data needs checking. 
Table~\ref{tab:datacheckcompare} shows the comparison between \cvi and the previous 
data integrity checking techniques.

\begin{table}[t]
    \small
    \centering
    \caption{Comparison between Different Data Integrity Mechanisms (R:Read, W:Write)}
    \label{tab:datacheckcompare}
    \vspace{-1ex}
    \begin{tabular}{ llll}
     \toprule
     Name & Instrumentation & Whitelist  &  Check \\ 
     \midrule
     DFI~\cite{Castro2006}        & All memory R\&W & Yes  &  Addr.\\
     WIT~\cite{akritidis2008preventing} & All memory W & Yes  &  Addr. \\
     DataShield~\cite{carr2017datashield} & All memory W & No  &  Addr. \\
     \cvi       & Critical variable R\&W   & No & Value \\
    \bottomrule
    \end{tabular}
\end{table}

\point{Critical Variable Identification \& Expansion}
\toolname compiler automatically identifies {\em control-dependent
variables}, including branch/loop condition variables. Note that we do not
include code pointers as critical variables. This is because all control-flow
violations, including those resulted from corrupted code pointers, are captured
by the control-flow part of \propshortname attestation. \toolname relies on
programmers to annotate {\em semantically critical variables} because
labeling such variables is subjective and specific to program
semantics.

The control-dependent variables (automatically detected) and the semantically
critical variables (annotated by programmers) form the initial set of critical
variables. Our compiler then automatically expands this set to include the
direct and indirect {\bf pointers} to these variables and their {\bf
dependencies}.

Pointers to a critical variable allow indirect define or use of
the variable. Therefore, we treat such data pointers as special critical
variables, referred to as {\em critical pointers}. Our compiler iteratively
identifies them from a global points-to map produced by the standard Anderson
pointer analysis.

Dependencies of a critical variable {\tt V} are the variables that may influence
the value of {\tt V}. Verifying the integrity of {\tt V} implies verifying the
integrity of both {\tt V} and its dependencies. Given a set of critical
variables, our compiler finds all their dependencies by first constructing the
program dependence graph and then performing a backward traversal for each
variable along the data dependence edges. Newly discovered variables during the
traversal are added to the critical variable set. The iterative search for
dependencies stops when the set reaches a fixpoint. 
This automatic dependency discovery also simplifies critical variable
annotation: programmers only need to annotate one, not all, variable for each
piece of critical data.

\point{\datacheckname}
\toolname compiler instruments all define- and use-sites for each variable in
the expanded critical variable set. During runtime, the instrumentation at each
define-site captures the identity of the critical variable (i.e., a
compiler-generated label) and the value to be assigned to the variable. It sends
the information to the measurement engine in the Secure World via the
trampolines. Similarly, the use-site instrumentation captures the variable
identity and the current value of the variable and sends them to the measurement
engine. 

For an array-typed critical variable, each array element access is instrumented
by \toolname compiler. \toolname compiler identifies and instruments every
memory access whose target address is calculated as an offset relative to a
critical variable. This critical variable can be an array name or a pointer that
points to a buffer. This design also uniformly covers 
multi-dimensional arrays and nested arrays because in both cases an array element
access is based on a memory address that is calculated from the array's name or
base address. From the measurement engine's perspective, 
it only sees 
critical variables or critical array elements identified by their 
addresses, rather than entire array objects. The integrity of each element is checked independently during runtime.

The measurement engine maintains a 
hashmap which stores pairs of {\tt <VariableID, Value>}. $VariableID$
is the address of a variable or an array element. 
The measurement engine updates the hashmap at each instrumented define-site and
checks the value at each instrumented use-site. Regardless whether a variable is
on the stack (local) or the heap (global), its def-use sites should always have
matching values. A value mismatch indicates a corrupted critical variable. The
measurement engine finally sends the \cvi checking result along with the
control-flow measurements in a signed blob to the remote verifier.

\point{Pointer-based Access to Critical Variables}
Consider an example: {\tt *(P+n) = x},
where {\tt P} is the base address of a critical array {\tt A}; {\tt n} is the
dynamic index. This is a legitimate critical variable define-site. Assuming
that, due to a program bug, the dereference {\tt *(P+n)} could go out of the
bounds of {\tt A} and reach another critical variable {\tt B}, the measurement
engine would then mistakenly update the value of {\tt B} to {\tt x} in its
hashmap when this out-of-bounds write happens.

We solve this issue by enforcing dynamic bounds checking on critical pointers.
When a critical pointer is dereferenced, the measurement engine checks if the
pointer dereference breaches the bounds of the current pointee. This check
relies on the dynamic bounds information (similar to
SoftBound~\cite{nagarakatte2009softbound}) collected by the measurement engine
for each critical pointer. If a bounds breach is found, the measurement engine
performs \cvi check only on the overlapping bytes between the accessed memory
region and the initially pointed variable (i.e., it only updates or checks the
value of the expected pointee). This design ensures \cvi check correctness while
allowing intentional out-of-bounds pointer dereferences in normal programs.

\section{Implementation}
\label{sec:impl}

Our \toolname prototype implementation includes {\em 6,017} lines of C++ code
for the compiler (an LLVM module), {\em 440} lines of C and assembly code for
the trampoline library, {\em 476} lines of C code for the measurement engine,
and {\em 782} lines of Python code for the verification engine.

\point{Hardware \& Software Choices}
We selected a widely used IoT development board, the HiKey board (ARM
Cortex-A53), as our reference device for prototyping. We made this choice due to the
board's unlocked/programmable TrustZone, its full compatibility with open-source
TEE OS, and its reliable debugging tools. We do realize that the board has a
relatively powerful processor compared to some low-end embedded devices.
However, no development board currently available comes with a low-end ARM
processor and an unlocked TrustZone at the same time. Nevertheless, \toolname's design and
implementation are neither specific to this board nor dependent on powerful
processors. The only hardware requirement \toolname has is the TrustZone
extension, which is available on many commercial embedded ARM SoCs.

We used OP-TEE~\cite{opteeurl} as our TEE OS (the OS for the Secure World).
Although \toolname is designed for bare-metal embedded devices (i.e., no
stand-alone OS in the Normal World), we used the
vendor-customized Linux as the Normal World OS solely for the purpose of easily booting
and debugging the board. \toolname itself does not assume the presence of a
Normal World OS. 
Even though \toolname's performance can be
negatively affected by the unnecessary OS, our evaluation (\S\ref{sec:eval})
shows the overhead under this disadvantaged configuration is still acceptable.

The implementation details of the \toolname system is attached in Appendix \S\ref{sec:impl_apx}.

\section{Discussion}
\label{sec:dis}

\point{Multithreading}
Our current prototype only supports single-thread programs, which constitute the
majority of today's embedded programs. To attest multi-thread programs, we need
to augment the \toolname compiler to instrument threading-related events and
have the measurement engine collect measurements and perform checks on a
per-thread basis. We consider multithreading support out-of-scope for this paper
because it does not need any change to the design of \propshortname attestation
but requires significant engineering efforts to implement. 

\point{Interrupt Handling}
Interrupt handling poses a challenge to the control flow verification due to its
asynchronous nature. 
When an interrupt happens in the middle of an attested operation, we do not know in
advance where the interrupted location is. If the measurement engine cannot recognize and process interrupts, they may introduce
out-of-context control flow events that can fail or confuse the verification. 

\toolname overcomes this challenge by treating each interrupt handler invoked
during an operation as an execution of a sub-program. It instruments the handler 
both at the entry and the exit points. The instrumentation notifies the measurement engine of the beginning and the end of such a sub-program.
Thus the control flow events of the handler
are recorded in a separate trace and hash and verified independently. \toolname also
checks if an invoked interrupt handler matches the interrupt that triggered the
handler.

\point{Annotation} \toolname allows programmers to optionally annotate
semantically critical variables for \cvi attestation. However, this annotation
is {\em not required} for detecting data-oriented programming~\cite{hu2016data},
control-flow bending~\cite{carlini2015control}, or similar data-only attacks,
whose target data (i.e., control-dependent variables) are automatically detected
by \toolname compiler and included in \cvi verification. On the other hand, the
annotation process for semantically critical variables is simple and facilitated
by the compiler's automatic dependency analysis.

\section{Evaluation and Analysis}
\label{sec:eval}

We conducted: (\rom{1}) micro performance tests to measure the overhead of
 each step in \propshortname attestation; (\rom{2}) macro performance tests on
5 real embedded programs of different kinds to examine \toolname's overall
overhead; (\rom{3}) tests against possible attacks; (\rom{4}) analysis on how \toolname
defends against evasions. 

\subsection{Micro Performance Tests}

The runtime overhead of \propshortname attestation can be broken down into three
parts, as shown in the first column in Table~\ref{tab:rtOverheadBreakdown}:
(\rom{1}) attestation initialization ($O_{at\_init}$), taking place at the entry
of an attestation scope (i.e., an operation); (\rom{2}) trampoline invocation
($O_{tramp}$), including a direct {\tt smc} call to initiate the world switch
and a return from the Secure World (i.e., the roundtrip world-switch overhead); 
(\rom{3}) attestation exit ($O_{at\_exit}$), happening at the end of an
attestation scope.

We created a test program that incurs each type of overhead exactly once. We ran
this program for 1,024 times and obtained the average overhead in terms of CPU
cycles as well as time used, as shown in Table~\ref{tab:rtOverheadBreakdown}.
$O_{at\_init}$ and $O_{at\_exit}$ are orders of magnitudes larger than
$O_{tramp}$. This is because they involve establishing or terminating the
attestation session and the communication between the Normal and the Secure
worlds. 
Since the initialization and exit happen only once for each attestation, their
overhead tends to blend in the (much longer) operation execution time and is
unnoticeable.

\begin{table}[]
    \small
    \centering
    \caption{Runtime Overhead Breakdown}
    \label{tab:rtOverheadBreakdown}
    \begin{tabular}{ llcc }
     \toprule
      \multicolumn{2}{l}{Overhead Sources} &  Time(CPU Cycles) & Time(ms) \\
     \midrule
     \multicolumn{2}{l}{Attestation Init ($O_{at\_init}$)}  & 5.1*$10^7 $ & 42.879\\
     \multicolumn{2}{l}{Trampoline Func($O_{tramp}$)} & 5.5*$10^4$ & 0.045 \\
     \multicolumn{2}{l}{Attestation Exit ($O_{at\_exit}$)} & 2.6*$10^7$ & 21.351\\
    \bottomrule
    \end{tabular}
\end{table}

\subsection{Tests on Real Embedded Programs}
\label{sec:evalreal}

\begin{savenotes}
    \begin{table*}[t]
        \small
        \centering
        \caption{Runtime overhead measured on 5 real embedded programs}
        \label{tab:runtime}
        \makebox[\textwidth][c]{
            \begin{tabular}{ l|ccc|ccccc|c|c}
                \hline
                \multirow{2}{*}{Prog.} & \multicolumn{3}{c|}{Operation Exec. Time} &\multicolumn{5}{c|}{\toolname Instrumentation Statistics} &
                Blob & Verification \\
                & w/o \propshortname (s) & w/ \propshortname (s) & Overhead (\%) & B.Cond & Def-Use & Ret & Icall/Ijmp & Critical Var. &  Size (B) & Time (s)\\
                \hline
                {\tt SP}  & 10.19 & 10.38  & 1.9\% & 488 & 2 & 1946& 1 & 20 & 69& 5.6 \\
                {\tt HA}  & 5.28  & 5.36   & 1.6\% & 147 & 91& 33  & 2 & 6  & 44& 0.61\\
                {\tt RM}  & 10.01  & 10.13  & 1.3\%  & 901 &100&100 & 100 & 7 & 913& 1.74\\
                {\tt RC}  & 2.55  & 2.66   & 4.5\% & 14  & 33& 1   & 1 & 8  & 10 & 0.13\\
                {\tt LC}  & 5.33  & 5.56   & 4.4\% & 931 &2420& 10 & 10 & 4  & 205& 1.35\\
                \hline
                Avg.  & N/A   & N/A    & 2.7\% & 496 & 529& 418& 23 & 9  & 248  & 1.89\\
                \hline
            \end{tabular}
        }
    \end{table*}
\end{savenotes}

We selected 5 open-source embedded programs to evaluate the end-to-end overhead
of \toolname. We could not test more programs because of the non-trivial manual
effort required for porting each program to our development board. This
requirement is due to embedded programs' device-specific nature. It is not posed
by our system. We picked these programs because they represent a reasonable
level of variety. Some of them may seem toy-like, which is not intentional but
reflects the fact that most embedded programs are simple by design. We note that
these programs are by no means standard benchmarks. In fact, there are no
standard benchmarks for bare-metal embedded devices available at the time of
writing. It is also rare for embedded device vendors to publicly release their
firmware in source code form. The 5 selected embedded programs are: 

\begin{itemize}
    \item \textit{Syringe Pump} ({\tt SP}) is a remotely controlled liquid-injection device, often used in healthcare and food
    processing settings. We apply \propshortname attestation
    to the  ``injection-upon-command'' operation.
    
    \item \textit{House Alarm System} ({\tt HA})~\cite{alarm4pi} is an IoT device that, when user-specified conditions are met, takes a picture and triggers an alarm. 
    We apply
    \propshortname attestation to its ``check-then-alarm'' operation. 
    
    \item \textit{Remote Movement Controller} ({\tt RM})~\cite{rmcontroller} is an embedded device that allows the
    physical movement of its host to be controlled remotely, similar to the example given in ~\S\ref{sec:runningexample}.
    We attest the ``receive-execute-command'' operation.
    
    \item \textit{Rover Controller} ({\tt RC}) ~\cite{rccontroller} 
    controls the motor on a rover. We attest the ``receive-execute-command''
    operation. 
    
    \item \textit{Light Controller} ({\tt LC})~\cite{lightcontroller} is a smart lighting controller. We attest the ``turn-on/off-upon-command'' operation.
\end{itemize}

\begin{figure}
    \centering
      \includegraphics[scale=0.45]{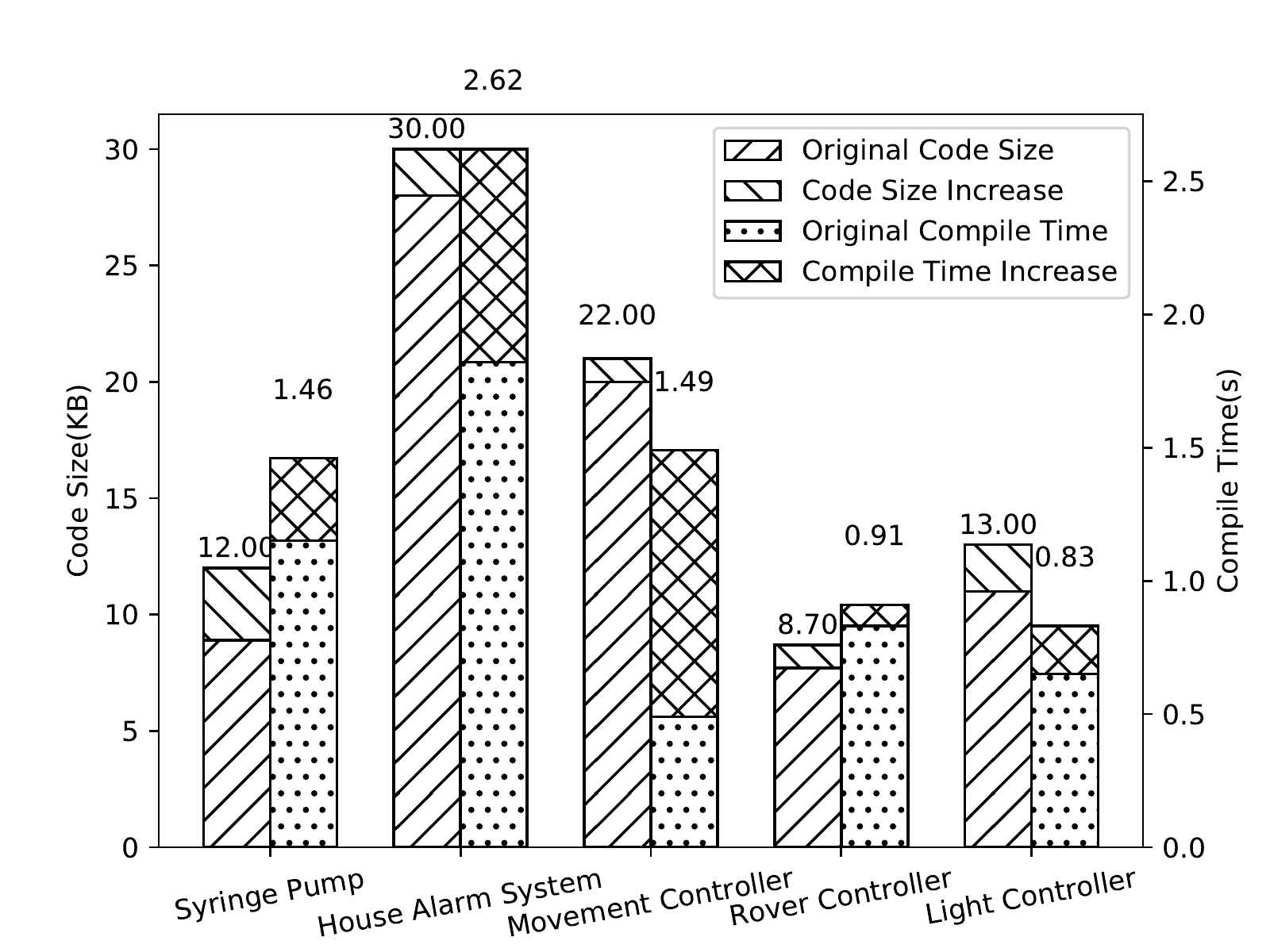}
      \caption{Compile-time overhead}
      \label{fig:static_overhead}
    \end{figure}

\point{Compile-time Overhead} We instrumented \toolname compiler to measure its
own performance in terms of binary size increase and compilation delay.
Figure~\ref{fig:static_overhead} shows the results for each program. The
absolute increase in code size ranges from 1 to 3 {\tt KB}, which is acceptable
even for embedded devices, despite the seemingly large percentage (13\%). We
find that the size increase is not proportional to the original code size and is
dominated by the trampoline library, which has a fixed size.

The compilation delay caused by the attestation-related code analysis and
instrumentation may seem high, averaging 62\%. But in absolute terms, the delay
is less than 1 second for the tested programs. We believe this is acceptable
given that (\rom{1}) the overhead is on par with similar code analysis and
instrumentation techniques; (\rom{2}) the compilation is offline and happens
only once for each program version.

\point{Operation Execution Time \& Instrumentation Statistics} For each test
program, we measured the execution times with and without OEI attestation 
enabled for the selected operation. The results are shown in the column named
``Operation Exec. Time'' in Table~\ref{tab:runtime}. The sub-column ``Overhead''
shows the relative delay caused by \toolname to operation executions, averaging
2.7\%. We observed that the delays are unnoticeable and blend in the much longer
end-to-end execution time of the operations.   

It is worth noting that the execution delay caused by our attestation may vary
significantly as the following two factors change: the length/duration of the
attested operation and the frequency of critical variable def-use events. For shorter
operations, the attestation overhead tends to be higher percentage-wise. For
instance, the operation in {\tt RC} (the shortest among the tested programs)
takes 2.55 seconds to finish without attestation and 2.66 seconds with
attestation, resulting in the highest relative overhead (4.5\%) among all tested
programs. However, this does not mean the absolute delay, in this case, is longer
than others or unacceptable.

The more frequent the def-use events of critical variables are, the higher the attestation delay
becomes. For example, the operations in {\tt HA} and {\tt LC} are similar in lengths. 
But the attestation overhead on {\tt HA} (1.3\%) is lower than the
overhead on {\tt LC} (4.4\%) partly because {\tt HA} has fewer def-use events of
critical variables. 

In the ``Instrumentation Statistics'' column, we show,  
for each operation execution, the numbers of the instrumented events encountered 
during the attestation (including conditional branches, def-use checks, returns, 
and indirect calls/jumps) as well as the number of critical variables selected. 
These statistics provide some insights into the overhead reported earlier. 
The instrumented events occur about thousands of times during an operation, 
which translates roughly to an average per-operation delay of 0.15 seconds. 

\point{Measurement Engine Memory Footprint and Runtime Overhead}
The measurement engine inside TEE consumes memory mainly for three purposes: the BLAKE-2s HASH calculation, the critical data
define and use check, and the forward control-flow trace recording, including taken or
not-taken bits and indirect jump/call destination. The BlAKE-2 HASH function 
only requires less than 2KB for storing the block buffer, 32 Bytes for the IV, 
160 Bytes for the sigma array, and some temporary buffers. The critical data
check requires a static HASH table of 4KB with 512 slots, and a dynamic
pool for critical variables (the size of this pool is proportional to the number
of critical variables; in our evaluation, the pool size is less than 2KB). The forward control-flow 
trace in our evaluation is no more than 2KB. The whole memory
footprint of the measurement engine is less than 10KB for the 
real embedded applications used in our evaluation.  

The runtime overhead of the measurement engine comes from three sources (i.e.,
the three major tasks of the measurement engine): calculating the hash upon
function return events, recording critical data define events, and verifying
critical data use event. In our evaluation, on average,  processing one return
event and calculating the new hash takes 0.19\us; recording a critical data
define event takes 11.04\us; verifying a critical data use event takes 2.03\us.
Obviously, hash calculation is relatively fast whereas critical data event
processing requires a longer time mainly because it involves hash table lookup or
memory allocation for a new entry.

\point{Value-based Check vs. Addressed-based Check}
To show the performance difference between value-based checking (\cvi) and
address-based checking (e.g., DFI), we measured the number of instrumented
instructions needed in both cases for all of the test programs. As shown in
Table~\ref{tab:comp_instrum}, on average, \cvi's instrumentation is 74\% less
than the instrumentation required by address-based checking (i.e., a 74\%
reduction). Specifically, \cvi's instrumentation is as little as 6.8\% of what
DFI requires when the program is relatively large (e.g., {\tt RM}). The number
only increases to 44.4\% when we annotated most of the variables as critical
in {\tt RC}.

\begin{table}[t]
	\small
	\centering
	\caption{Number of Instrumentation Sites: Value-based (R1) and Address-based (R2)}
	\label{tab:comp_instrum}
	\vspace{-1ex}
	\begin{tabular}{ l|cccccc}
		\toprule
		              & SP   & HA    & RM    & RC     & LC    & Avg.\\
		\hline
		R1   & 56   & 37    & 57    & 20     & 41    & - \\
		R2 & 140  & 388   & 842   & 45     & 131   & - \\
		R1 / R2    & 40\% & 9.5\% & 6.8\% & 44.4\% & 31.2\% & 26\% \\
		\bottomrule
	\end{tabular}
\end{table}

\point{Space-efficiency of Hybrid Attestation}
Our control-flow attestation uses the hybrid scheme consisting both forward
traces and backward hashes to achieve not only complete verifiability but also
space-efficiency. To quantify the space-efficiency, we compared the sizes of the
control-flow traces produced by \toolname (R1 in Table~\ref{tab:comp_trace}) and
the traces produced by pure trace-based CFI (R2 in Table~\ref{tab:comp_trace}).
On average, \toolname's traces take only 2.24\% of space as needed by
control-flow
traces (i.e., a 97\% reduction). This result shows that our hybrid scheme is
much better suited for embedded devices than solely trace-based CFI in terms of
space efficiency.

\begin{table}[t]
	\small
	\centering
	\caption{Control-flow Trace Size (Bytes): With Return Hash (R1) and Without Return Hash (R2)}
	\label{tab:comp_trace}
	\vspace{-1ex}
	\begin{tabular}{ l|cccccc}
		\toprule
		             & SP    & HA     & RM     & RC      & LC      & Avg.\\
		\hline
		R1           & 69    & 44     & 913    & 10       &  205    & - \\
		R2           & 42941 & 3772   & 13713  & 585     & 13725   & - \\
		R1 / R2      & 0.2\% & 1.1\%  & 6.7\%  & 1.7\%   & 1.5\%    & 2.24\% \\
		\bottomrule
	\end{tabular}
\end{table}

On the other hand, compared with existing hash-based attestation schemes,
\toolname's attestation blobs are not of fixed-length and grow as attested
operations execute, which may lead to overly large attestation blobs. 
However, in
practice, \toolname attestation blobs are reasonably small in size. 
Based on our experiments,
the average blob size is $0.25 kb$. 
The individual blob size for each program is shown in the ``Blob Size'' column in
Table~\ref{tab:runtime}. We attribute this optimal result to two design choices: (\rom{1}) the
hybrid measurement scheme that uses fixed length hashes for verifying returns (more frequent)
and traces for verifying indirect forward control transfers (less frequent);
(\rom{2}) the operation-scoped control-flow attestation, which generates
per-operation measurements and is enabled only when an operation is being
performed.

\point{Verification Time}
We also measured \toolname verifier's execution time when it checks the
attestation blob generated for each program. The result is shown in the
``Verification Time'' column in Table~\ref{tab:runtime}, averaging 1.89
seconds per operation. It shows that verification is not only deterministic but fast. 
On average, the verification is one order of magnitude faster than the original
execution (Table~\ref{tab:runtime}). This result echoes that the verification is
not a re-run of the program. It is a static abstract execution guided by the
measurement stream.

\subsection{Attack Detection via OEI Attestation}
\label{sec:security_analysis}
Due to the lack of publicly available exploits for bare-metal devices, we injected
vulnerabilities to the previously discussed test programs, launch basic
control-flow hijacks and data corruption, and examine if the measurements
generated by \toolname capture these attacks. 

Specifically, we injected to the programs the vulnerabilities similar to those
shown in Listing~\ref{lst:examplecode}. We then exploited the vulnerabilities to
(\rom{1}) overwrite a function pointer; (\rom{2}) corrupt a critical variable;
(\rom{3}) trigger an unintended operation. By verifying the measurements
generated by \toolname, we found that, in each test case, (\rom{1}) the illegal
control-flow transfer caused by the subverted function pointer is recorded in
the measurement stream; (\rom{2}) the \cvi flag is set due to the failed \cvi
check; (\rom{3}) the unintended operation is detected because the
reconstructed code path does not match the requested operation. 

Although these tests are simple and created by ourselves, they do demonstrate
the basic functioning of our prototype and confirm OEI attestation as a viable
way for remote verifiers to detect those attacks that are currently undetectable
on embedded devices. Moreover, they showcase that IoT backend can now use
\toolname to remotely attest the operations performed by IoT devices and
establish verifiable trust on these devices.

\subsection{Security Analysis}
\label{sec:security}
Our threat model (\S\ref{sec:threatModel}) anticipates that attackers may find
and exploit unknown vulnerabilities in the embedded programs running in the
Normal World. However, we assume code injection or modification cannot happen,
which the existing code integrity schemes for embedded devices already
prevent~\cite{clements2017protecting,kimsecuring}.

To evade \toolname, a normal-world attacker would need to \textcircled{1}
disable the instrumentation or the trampolines, \textcircled{2} abuse the
interfaces that the measurement engine exposed to the trampolines, or CA-TA
interfaces, \textcircled{3} manipulate the control flow in a way to generate a HASH
collision, thus bypassing the verification, or \textcircled{4} modify the
attestation blob including replay an old recorded blob.  

\textcircled{1} is ruled out by the code integrity assumption. Plus,
attempts to divert the control-flows of instrumented code or trampolines are
always recorded in the control-flow trace and detected later by the verifier.
Our design prevents \textcircled{2} as follows. \toolname compiler disallows
world-switching instructions ({\tt smc}) used outside of the trampoline library.
This restriction ensures that only the trampoline functions can directly invoke
the CA-TA interfaces and the rest of the code in the Normal World cannot. 
To further prevent code-reuse attacks (e.g., jumping to the interface invocation
point in the library from outside), \toolname loads the library in a designated
memory region. The compiler masks the target of every indirect control transfer
in the embedded program so that the trampoline library cannot be reached via
indirect calls or jumps or returns (i.e., only hard-coded direct calls from each
instrumentation site can reach trampolines). This masking-based technique is
highly efficient and is commonly used for software fault isolation. As a result,
\textcircled{2} is prevented. 

As for \textcircled{3}, we assume that the attacker may exploit program
vulnerabilities and manipulate the control flow of the program in arbitrary
ways. We prove that (see Appendix~\ref{sec:prf}) our control-flow verification
mechanism cannot be bypassed by such a powerful attacker. Our proof shows that
bypassing our verification is at least as hard as finding a hash collision, which is
practically infeasible considering that BLAKE-2s is as collision-resistent as
SHA3~\cite{blake2}. 

Our verification scheme prevents \textcircled{4} because the integrity of the
attestation blob is guarded by a signature generated from TEE with a
hardware-provisioned private key.  A verifier can easily check the signature and
verifies the integrity of the attestation blob. Replay attack is also prevented
by checking whether the cryptographic nonce inside the attestation blob matches
what originally was generated by the verifier.

There is no higher privileged code (e.g., a standalone OS) that needs to be
protected or trusted because \toolname targets bare-metal embedded devices. 
For the same reason, it is realistic to require the firmware to be
entirely built using \toolname compiler. 

\section{Related Work}
\point{Remote Attestation}
Early works on remote attestation, such as \cite{Arvind04softAt}\cite{Li2011},
were focused on static code integrity, checking if code running on remote
devices has been modified. A series of works~\cite{Arbaugh1997,Parno2010,
Eldefrawy2012,Kong2014} studied the Root of Trust for remote attestation,
relying on either software-based TCB or hardware-based TPM or PUF. Armknecht et
al.~\cite{Armknecht2013} built a security framework for software attestation. 

Other works went beyond static property attestation. Haldar et
al.~\cite{Haldar04semantic} proposed the verification of some high-level
semantic properties for Java programs via an instrumented Java virtual machine.
ReDAS~\cite{Kil09ReDAS} verified the dynamic system properties. Compared with
our work, these previous systems were not designed to verify control-flow or
dynamic data integrity. Further, their designs do not consider bare-metal
embedded devices or IoT devices. Some recent remote attestation systems
addressed other challenges. A tool called DARPA~\cite{Ibrahim2016} is resilient
to physical attacks. SEDA~\cite{Asokan2015} proposed a swarm attestation scheme
scalable to a large group of devices. In contrast, we propose a new remote
attestation scheme to solve a different and open problem: IoT backend's
inability to verify if IoT devices faithfully perform operations without being
manipulated by advanced attacks (i.e., control-flow hijacks or data-only
attacks). Our attestation centers around \propshortname, a new security
property we formulated for bare-metal embedded devices. \propshortname is
operation-oriented and entails both control-flow and critical data integrity.

A recent work called C-FLAT~\cite{Abera2016} is closely related to our work. 
It enabled control-flow attestation for embedded devices. However, it suffers
from unverifiable hashes, especially when attested programs have nested loops
and branches. This is because verifying a control-flow hash produced by C-FLAT
requires the knowledge of all legitimate control-flow hashes, which are
impossible to completely pre-compute due to the unbounded number of code paths
in regular programs (i.e., the path explosion problem). In comparison, \toolname
uses a new hybrid scheme for attesting control-flows, which allows deterministic
and fast verification. 
Moreover, \toolname verifies \cvifull and can detect data-only attacks, which
C-FLAT and other previous works cannot.

\point{Online CFI Enforcement}
Although control-flow attestation has not been well investigated, CFI
enforcement is a topic that has attracted a rich body of works since its
debut~\cite{abadi2005control}. 
A common goal shared by many CFI enforcement methods such as
\cite{tice2014enforcing,Niu2015,zeng2011combining,niu2013monitor,criswell2014kcofi,davi2012mocfi} 
is to find a practical trade-off between runtime overhead and the level of
precision. 
Previous works such as~\cite{zhang2013control,zhang2013practical} also
introduced CFI enforcement to COTS or legacy binaries. CPI~\cite{Kuznetsov2014}
prevents control flow hijacking by protecting code pointers in safe regions.
These work made CFI increasingly practical for adoption
in the real world and serves as an effective software exploitation prevention
mechanism. 

However, enforcing fine-grained CFI and backward-edge integrity can be still too
heavy or impractical for embedded devices, mostly because of the limited CPU
and storage on such devices. 
Apart from less demanding on hardware resources, OEI attestation has another
advantage over online CFI enforcement: it allows remote verifiers to reconstruct
the exact code paths executed during an operation, which enables full CFI
checking (as opposed to a reduced or coarse-grained version) as well as other
postmortem security analysis.

Moreover,  CFI enforcement is not enough when it comes to data-only attacks, such
as control-flow bending, data-oriented programming,
etc.~\cite{Chen2005,carlini2015control,hu2016data,Hu2015}. But these attacks do
violate \propshortname and can be detected by \toolname.

\point{Runtime Data Protection}
A series of work addressed the problem of program data corruption via
dynamic bounds
checking~\cite{dhurjati2006backwards,nagarakatte2009softbound,devietti2008hardbound}
and temporal safety~\cite{nagarakatte2010cets}. DFI~\cite{Castro2006} and WIT~\cite{akritidis2008preventing} took a different approach. They use static
analysis to derive a policy table specifying which memory addresses each instruction
can write to. They instrument all memory-access instructions to ensure the policy is
not violated during runtime. Although effective at preventing data corruption,
these techniques tend to incur high runtime overhead due to the need to
intercept and check a large number of memory accesses. We refer to this line of work as address-based checking. 
In contrast, we define \cvifull and use the new \datacheckname to verify \cvi.
Our check is selective (i.e., it only applies to critical variables) and
lightweight. It is value-based and does not require complex policies or extensive
instrumentation. The \cvi verification and the control-flow attestation mutually
compensate each other, forming the basis for \propshortname verification. 

DataShield\cite{carr2017datashield} applies selective protection to sensitive
data. Their definition of sensitive data is type-based and also needs programmer
annotation. It relies on a protected memory region to isolate the sensitive data
and performs address-based checking. In comparison, our critical variable
annotation is more flexible and partly automated. Instead of creating designated
safe memory regions, which can be unaffordable or unsupported on embedded
devices, we perform lightweight value-based checks. Unlike DataShield,
\propshortname does not concern data confidentiality.

\section{Conclusion}
We tackle the open problem that IoT backends are unable to remotely detect
advanced attacks targeting IoT devices. These attacks compromise control-flow or
critical data of a device, and in turn, manipulate IoT backends. We propose
\propshortname, a new security property for remotely attesting the integrity of
operations performed by embedded devices. \propshortname entails {\em
operation-scoped CFI} and {\em \cvifull}. 

We present an end-to-end system called \toolname that realizes \propshortname
attestation on ARM-based bare-metal embedded devices. \toolname solves two
research challenges associated with attesting dynamic control and data
properties: incomplete verification of CFI and heavy data integrity checking.
First, \toolname combines forward-edge traces and backward-edge hashes as
control-flow measurements. It allows fast and complete control-flow verification
and reconstruction while keeping the measurements compact. Second, \toolname
enforces selective value-based variable integrity checking. The mechanism is
lightweight thanks to the significantly reduced instrumentation. It enables the
detection of data-only attacks for the first time on embedded devices. 
It allows IoT backends to establish trust on incoming data and requests from IoT
devices.

\bibliographystyle{plain}
\bibliography{bibliography}

\begin{appendices}
\section{Implementation Details of \toolname System}
\label{sec:impl_apx}

\point{Compiler-based Instrumentation}
\toolname compiler is built on LLVM~\cite{LLVM:CGO04}. Besides the typical
compilation tasks, it performs (\rom{1}) the analysis for identifying critical
variables;  (\rom{2}) the code instrumentation for collecting runtime
measurements on control flow and critical variables. 
The analysis works on the LLVM IR. It first constructs the initial set of critical
variables by traversing the IR and searching for control-dependent variables and
programmer-annotated semantically critical variables. It then uses a
field-sensitive context-insensitive Anderson pointer analysis to generate the
global points-to information. The compiler uses this point-to information to
recursively identify direct and indirect pointers to the critical variables
(i.e., critical pointers). It also performs a backward slicing for each critical
variable on the program dependence graph to find its dependencies. All critical
pointers and dependencies are iteratively added to the critical variable set. 

\toolname compiler instruments the code via an added backend pass in LLVM. The
instrumentation is needed at both the assembly level (for control-related
instructions) and the IR level (for data-related instructions). This is
important because the translation from the IR to the machine code can generate
additional control-flow instructions that need to be instrumented. The compiler
inserts calls to trampolines at instructions that can change the control flow of
attested operations or store/load critical variables
(Table~\ref{tab:instrumentation}). Though seemingly straightforward, this
instrumentation, if not designed carefully, can break the original program
because a trampoline call may corrupt the destination register used by the
original control transfer.  To avoid such issues, the instrumentation saves to
the stack the registers that are to be used for passing the parameters to the
trampoline. Moreover, the trampoline is responsible for handling the
caller-saved registers (normally handled by callers rather than callees). This
design reduces the number of inserted instructions at each instrumentation site.
It also minimizes the stack frame growth. As a result, registers changed during a
trampoline call are restored immediately after the call returns.       

\begin{table}[t]
\small
\centering
\caption{Instrumented Instructions for Control and Data Measurement}
\label{tab:instrumentation}
\begin{tabular}{ llll }
 \toprule
 Inst Type& Layer &Inst & Info to Record\\ 
 \hline
 Ind. Call & Assm. & \texttt{blr xr} & \texttt{xr} \\ 
 Ind. Jump& Assm. & \texttt{br xr} & \texttt{xr} \\  
 \multirow{2}{*}{Cond. Jump} & \multirow{2}{*}{Assm.}& \texttt{b.cond,cbz} & \multirow{2}{*}{\texttt{true/false}} \\  
 & & \texttt{cbnz,tbz,tbnz} & \\
 Data Access& IR& \texttt{load/store} & \texttt{addr,value} \\  
 Return & Assm. &\texttt{ret} & \texttt{pc,lr}\\ 
 \bottomrule
\end{tabular}
\end{table}

\point{Measurement Engine}
We built the measurement engine as a Trusted Application (TA) running in the
Secure World. It handles events generated by the trampolines (i.e., the Client
Application, or CA) during runtime. Control-flow events are only generated and
handled during an active attestation window (when an attestation-enabled
operation is executing). Internally, the measurement engine maintains, for each
active operation, a {\em binary trace} ($S_{bin}$) for branches, an {\em
address trace} ($S_{addr}$) for indirect calls/jumps, and a {\em hash} ($H$)
for returns. 
At the end of an attestation session, the engine concatenates $S_{bin}$ and
$S_{addr}$ to form a single measurement stream, $S$, in a sequentially parsable
format: $Size(S_{addr})|S_{addr}|Size(S_{bin})|S_{bin}$. 

Data load/store events are only triggered by critical variables. To perform \cvi
check, the engine maintains a  hashmap to keep track of each critical
variable's last-defined value. At every use-site of a critical variable, the
engine checks if the observed value equals the stored value in the hashmap.
If a mismatch is encountered, the engine sets a global flag, $F$, to indicate
the \cvi violation. If a violation is detected, the engine also records the variable address and the
previous return address as the context information $C$, which allows the remote
verifier to investigate the violation. Finally, the engine generates a signed
attestation blob that consists of $S$, $H$, $F$, and $C$ if \cvi verification
failed, along with a nonce $N$ sent
from the verifier who initiated the attestation. It will be passed back 
to the normal world who will finally send the signed attestation
blob to our verification engine
via the network. Although we use the normal world's network stack, 
we do not need to trust it. Any corruption of the blob is detectable by verifying the signature. Any
denial of service by the normal world network stack also causes attestation failure.

\point{CA-TA Interaction}
We implemented three CA-TA communication interfaces compliant with the
GlobalPlatform's TEE specification~\cite{gpspec}, a de-facto standard for
TEE interface design. The interfaces are {\tt oei\_attest\_begin}, {\tt
commit\_event}, and {\tt oei\_attest\_end}, used by the CA to notify the TA of
the respective event. To prevent potential abuse (e.g., calling them via ROP),
the measurement engine ensures that the interfaces can only be called by the
trampolines and can never be invoked via indirect calls, jumps, or returns
(details in \S\ref{sec:security} ).

\point{Verification Engine}  
We prototyped a simple verification engine based on the Capstone
disassembler~\cite{capstone}. It takes as input an attestation blob, the
binary code that performed the operation under attestation, and a CFG extracted at
compile time for that operation code. As
described in ~\S\ref{sec:cf}, the verification process is fairly
straightforward, thanks to our hybrid measurement scheme.  

For control-flow attestation, the verifier performs a static abstract execution
of the disassembled binary code. This abstract execution is guided by the
forward-edge traces in the attestation blob. It simply traverses through the
code and performs CFI checks at each indirect control-flow transfer. It also
simulates a call stack for keeping track of return addresses and updating the
return hash, which is checked against the reported hash in the end. This
abstract-execution-based verification is fast because it does not actually run
the code or have to exhaustively explore all possible code paths. Moreover,
unlike traditional attestation, which only gives a binary result, our
verification allows for the reconstruction of the execution traces, which are
valuable to postmortem analysis. 
For \cvi verification, the verifier checks if the \cvi violation bit is
set in the attestation blog. If positive, it fails the attestation and outputs
the context information.

\section{Proof of Control-flow Verification}
\label{sec:prf}

Let $h : \{ 0,1 \}^\star \rightarrow \{ 0,1 \}$ be a collision
resistant hash function. Using $h$ we can construct another hash function
$H[op]$ which takes the sequence of values $\langle z_1,\cdots, z_m \rangle$
as follows: $H_1 = h(z_1)$ and $H_{i+1} = op (H_i,z_{i+1})$ ($1 \leq i
< m$).  The value of $H$ on the sequence is $H_m$. We rely on the
following property, which restricts the binary operation $op$ we can
use.
\begin{quote}
  If $h$ is collision resistant, then $H[op]$ is collision resistant
\end{quote}
Our implementation uses BLAKE-2s as $h$ and supports a variety of binary
operations, such as concatenation and xor. Recall that if $op$ is
concatenation, $H[op]$ is very similar to the classic Merkle-Damgrad
construction. For the rest of the note, we will fix the binary operation $op$ (\ie we use concatenation as $op$ in our implementation)
and just write $H$ instead of $H[op]$.

Let $P$ be the program under consideration.  Let $C(P) =
\{c_1,\cdots,c_k \}$ and $R(P) = \{ r_1,\cdots,r_k \}$ be the call and
return sites in a program $P$ (we will assume that the return site
$r_i$ corresponds to the call site $c_i$). Recall that a proof
$\sigma$ has three components $(\alpha,v,\beta)$, where $\alpha \in
C(P)^*$ (a sequence of call sites), $v$ is the hash value of the
sequence of returns, and $\beta$ is a sequence of jumps (direct and
indirect) and conditional branches (essentially $\beta$ has everything
related to control-flow transfers, except calls and returns). A path $\pi$
through the control-flow graph (CFG) program $P$ is called {\it legal}
if it satisfies two conditions: {\sf (A)} the call and returns in
$\pi$ are balanced~\footnote{This means that call and returns satisfy
  the grammar with the following rules: $S \rightarrow c_i \; S \;
  r_i$ (for $1 \leq i \leq k$) and $S \rightarrow \epsilon$.}, {\sf
  (B)} the jumps and conditional branches are legal (this can be
easily checked by looking at the source code of $P$ and the data
values corresponding to the targets of the indirect jumps). $\Pi (P)$
denotes the set of execution paths through the CFG of the program $P$. The
proof corresponding to an execution path $\pi \in \Pi(P)$ is denoted
by $\sigma (\pi)$. Next, we describe our verification algorithm.

\noindent
{\bf Verification.} Our verifier $vrfy (P,\sigma)$ 
(let $\sigma = (\alpha,v,\beta)$) and is conjunction of two verifiers
$vrfy_j$ and $vrfy_c$ described below.
\begin{itemize}
    \item Verifier $vrfy_j (P,\sigma)$ checks that the jumps and
      branches in $\beta$ are valid (i.e. this can be easily done
      because the verifier has the program $P$ and the data values
      corresponding to the targets of the indirect jumps).  If the
      jumps are valid, then $vrfy_j (P,\sigma) = 1$; otherwise $vrfy_j
      (P,\sigma) = 0$
    
    \item Verifier $vrfy_c (P,\sigma)$ checks the validity of calls
      and returns. This part is a bit more involved.  Essentially
      $vrfy_c$ ``mimics'' how the hash of the returns are computed and
      then checks if the computed hash value matches the one in the
      proof $\sigma$.  Verifier $vrfy_c$ maintains an auxiliary stack
      $st$ and processes the sequence of calls $\alpha = \langle
      c_{j_1}, \cdots, c_{j_n} \rangle$ as follows: The calls in
      $\alpha$ are processed in order, and the verifier keeps running
      hash. Let us say we have processed $c_{j_1},\cdots,c_{j_r}$ and
      are processing $c_{j_{r+1}}$. Recall that from the call site we
      can tell if there was a context switch in the program execution
      (a context switch means that we are executing in a different
      function). The call site has the location of the program, so we
      can inspect whether we are in the same function as the top of
      the stack (i.e., the location of $c_{j_{r+1}}$ is different from
      the location of the call site on top of the stack). If there was
      no context switch, then we push $c_{j_{r+1}}$ on the stack. If
      there was a context switch, then we pop the top of the stack
      (say $c_u$), compute $v' = op(v',h(r_u))$, and push
      $c_{j_{r+1}}$ on the stack.  If $r_u$ was the first return
      computed, then $v'=h(r_u)$. After all the calls have been
      processed, let the hash value be $v'$. The verifier $vrfy_c$
      outputs a $1$ if $v = v'$; otherwise it outputs a $0$.
\end{itemize}
The verifier $vrfy (P,\sigma)$ is 
$vrfy_j (P,\sigma) \wedge vrfy_c (P,\sigma)$.

\begin{definition}
A proof $\sigma$ is called {\it ambiguous} iff there are two paths
$\pi$ and $\pi'$ and $\pi$ such that: (I) $\sigma \; = \; \sigma(\pi)
\; = \; \sigma(\pi')$ (II) $\pi$ is legal and $\pi'$ is illegal.
\end{definition}
Note that if the verifier gets an ambiguous proof $\sigma$, then it
cannot reject it because it could also correspond to a legal path
$\pi$. Therefore, an adversary is free to take an illegal path $\pi'$
corresponding to the ambiguous proof.  Therefore, adversary's goal is
to generate an ambiguous proof.

Essentially the lemma given below informally states that {\it if an
  adversary can generate an ambiguous proof, then they can find a
  collision in the hash function $H[op]$}. Hence, if $H[op]$ is
collision resistant, then it will be hard for an adversary to find an
ambiguous proof and ``fool'' the verifier.

\begin{lemma}
If there exists an ambiguous proof $\sigma$, then there is a collision
in the hash function $H[op]$.
\end{lemma}
\noindent
{\bf Proof:} Let $\pi$ and $\pi'$ be two execution paths that result
in the same proof $\sigma$. Moreover, let $\pi$ be legal and $\pi'$ be
illegal (recall that $\sigma$ is an ambiguous proof). Let $r_\pi$ and
$r_{\pi'}$ be the sequence of returns for the two paths $\pi$ and
$\pi'$. The set of direct jumps and call sequences for the two paths
are the same (since they correspond to the same proof $\sigma$), so
the sequence of returns has to be different (otherwise the two paths
will be the same, which is a contradiction). However, the two
sequences of returns hash to the same value under $H[op]$ because the
paths correspond to the same proof. Thus, we have found a collision in
$H[op]$. $\Box$

\end{appendices}

\end{document}